# Mechanical properties of fat-based semi solid heterogeneous materials

**Henri de Cagny[1,2], Stefan Kooij[1], Peter Shall, Theo Blijdenstein[2], Luben Arnaudov[2], Simeon Stoyanov[2,3], Daniel Bonn[1].**

[1]*Van der Waals-Zeeman Instituut, Institute of Physics, University of Amsterdam, 1090 GL Amsterdam*
[2]*Unilever Research and development center, Olivier van Noortlaan 120, 3133 AT Vlaardingen*
[3]*Laboratory of Physical Chemistry and Colloid Science, Wageningen University, 6703 HB Wageningen, The Netherlands*

Fat based pasty products are an essential part of food industries, both as process intermediate or end products. Examples include chocolate spread, margarine, biscuits or bouillon cubes. Such materials are made of a continuous fat phase in which solid particles are dispersed. Phenomena such as yield and fracture occur in those materials and depend heavily on the composition of the samples. Because of its structure ranging from nano-to the macro scale (1) (2), crystallization and deformation of fat by itself is already a very dense topic, being the subject of dozens of studies (3) (4) (5) (6).

The crystallization of fat is altered in presence of particle, which leads to products with different physical properties. However, because of the very complex aspect of the problem, very few studies have been trying to characterize the influence of the solid particles on the final product (7) (8) (9). In these studies, small amplitude oscillatory rheology is often used to characterize the stiffness of the materials. But processed products might not meet the desired shape necessary to perform rheology experiments:  such measurements require a very specific product shape (cylindrical in the best case, conic for cone-plates experiments) that might not be easily achievable when it comes to industrial trials. Furthermore, oscillatory rheology also demands a complete absence of wall slip, a condition which can be hard to meet for semi solid samples. One could try to glue the material to the bottom and top part of the rheometer, but if the samples are slightly porous, the glue might pollute the experiment and this would result in completely wrong measurements. To perform quick and reliable experiments, oscillatory rheology is therefore not always the best option to measure the physical properties of hard fat mixes.

Micro indentation is an alternative to oscillatory measurements where one of the only requirements for the samples is to present a flat surface to the indentation probe. Such a measurement can usually be performed in the industry through the help of a texture analyzer. The theory behind the micro indentation of purely elastic solids has been derived by Hertz in (10), but theoretical studies deriving the relation between indentation, probe shape and complex modulus for visco-elastic materials started to appear not that long ago (11) (12) (13)

In this paper, we formulate binary mixes consisting of a hard fat matrix (made of palm oil stearin) initially melted, and in which we added solid inclusions of sodium chloride NaCl (sometimes referred as salt in the rest of the paper). We create various set of samples with different salt and fat volume fraction and let them crystallize for a few weeks. We then perform micro indentation protocols on the resulting and use two different protocols:

1) The Oliver-Pharr method (14) (15), where a probe initially indents the material, and where the unloading stage is also recorded. The difference between the loading and unloading curves can tell how reversible (or not) the deformations are
2) The stress relaxation method, where the indentation reaches very quickly a given depth, and where the relaxation of the force over time is recorded.

From these results, we then set up a small rheological model to derive the complex modulus of the product and try to predict the force vs deformation curves of those samples under other deformation protocol, checking if our model is consistent with other set of experiments. We show that when gradually replacing the fat matrix by salt particle, G* becomes high and higher, and the contribution from the elastic deformation dominates more and more compared to the viscous deformations and diverges at a given solid volume fraction.

# 2) Materials and methods

## 2.1) Samples

The materials all consist of heterogeneous mixes with a continuous phase (the fat) with solid inclusion (salt particles) trapped into it. Palm oil stearin, POs53, was used for the fat continuous phase. The stearin was purchased from Sime d'Arby Unimills B.V. (Zwijndrecht, the Netherlands) and was separated from the palm oil by (fractionation. The fat was characterized by a melting trajectory, indicated as an N-line which give the weight fraction of solid fat (N) at a given temperature. For this specific fat, $N_{10}$ = 85%, $N_{20}$ = 69%, $N_{30}$ = 45%, $N_{40}$ = 28%. The salt used was micro salt obtained from Akzo Nobel (Deventer, the Netherlands). The average size of the micro salt was characterized with a Morphologi G3S and the average size measured was 50$\mu$m. The precise composition of the fat and the size distribution of the salt are available in the supplementary information (SI). To prepare heterogeneous samples with the desired volume fraction, it is necessary to know the density of the fat phase. The density was measured with a densitometer DMA 45 at various temperature. At ambient temperature, the density is found to be equal to 0.939 kg.m$^{-3}$. The samples are then prepared by melting the fat phase at 90 °C. The fat is then poured in a Thermomix Food Processor and gently stirred while cooling down. Salt particles were also heated beforehand heated to 90°C so that all samples have the same cooling rate and are then added. The mixture is gently stirred to avoid sedimentation, until crystallization starts and a yield stress appears. The salt volume fraction $\phi_{NaCl}$ in the cubes varies between 0 and 60% (the rest of the sample being fat).

The slurry necessary to prepare a cube of dimension 23*28*11.5mm is then individually weighted for each cube. The resulting amount is then poured in a mold made of 20 rectangular holes of dimension (23*28mm) The slurry is then compressed by a universal testing machine (Instron 5567) until the samples reach the desired height (here 11mm). After compression, the cubes are stored 4 weeks at ambient temperature so that the crystallization process can finish. For each value of $\phi_{NaCl}$, two batches of 20 cubes are prepared.

## 2.2) Measurements methods

### 2.2.1) Density of the samples

To check the presence of air in the samples, the cubes were suspended in a solvent (ethanol) and the resulting change of weight measured. The density of the cubes can then be derived thanks to Archimedes' principle following the buoyancy method. The formula linking the density of the solid $\rho_s$, the density of the fluid $\rho_{fl}$, the mass of the solid in air $m_a$ and it's apparent mass when immersed $m_{fl}$ is given by :

$$\rho_s = \rho_{fl} \frac{m_a}{m_a - m_{fl}} \tag{1}$$

### 2.2.2) Micro indentation

Attempts to perform rheology measurements on our samples through classical oscillatory shearing were unsuccessful, for various reasons: oscillatory rheology measurements needs thin samples that can easily take a cylindrical shape, and can stick to the geometry to avoid wall slip (14) (15). Our samples had a cubic shape and are therefore unsuited to rotational measurements. We tried to chop our samples into a cylindrical shape with a cookie cutter, but the resulting structures were severely weakened, if not fragmented. Furthermore, because of the large height of our cubes, a small strain induced a large absolute displacement (16).Wall slip was already visible with the naked high at low strain. We also tried to glue the cube to the geometry, but the resulting moduli were a few orders of magnitudes harder than the usual results obtained for such fat systems: we ultimately ended partially measuring the flowing properties and modulus of the glue. We therefore had to use other methods to characterize the physical properties of our cubes.

Instrumented indentation is a very simple measurement where a probe with a defined size and shape indents a material and the resulting force is recorded (or vice versa). It is widely used in the study of small scale behavior of soft matters such as polymers, biomaterials or food products (17) (18) (19). One of the most famous theory for indentation between a probe and a sample is the Hertzian theory of non-adhesive contact (10), which describes the relationship between a semi-infinite elastic plane and an elastic probe with various geometrical shapes. But our samples show high non-reversible deformations which are not considered in such a simplistic model. Several studies have been published focusing on visco-elastic materials with different types of loading probes and protocols (12) (20) (13) (21). In this section, we will focus on the two main methods we have been testing in the paper: The Oliver-Pharr method and the relaxation response to a strain-controlled indentation. All our micro indentation measurements have been performed with a Texture Analyser (TA Instruments)

## 3)Results and Discussion

**3.1) Density measurements of the samples.**

Because of the low density of the fat phase, the samples float when immersed in water and we therefore need to use a less dense solvent to measure the density of the processed cubes, and therefore dip them in ethanol ($\rho = 789 kg/m^3$). The measures densities are shown on figure (1) and are compared with the theoretical density that one should expect in a sample without any air. As one can see, the densities are systematically lower than the expected one, a sign that some air has been trapped in the samples during the crystallization. When measuring the amount of air in the samples, the results were always comprised between 6 and 8 %, except for the samples with a $\phi_{NaCl} = 0.4$ which showed a higher amount of air (12%)

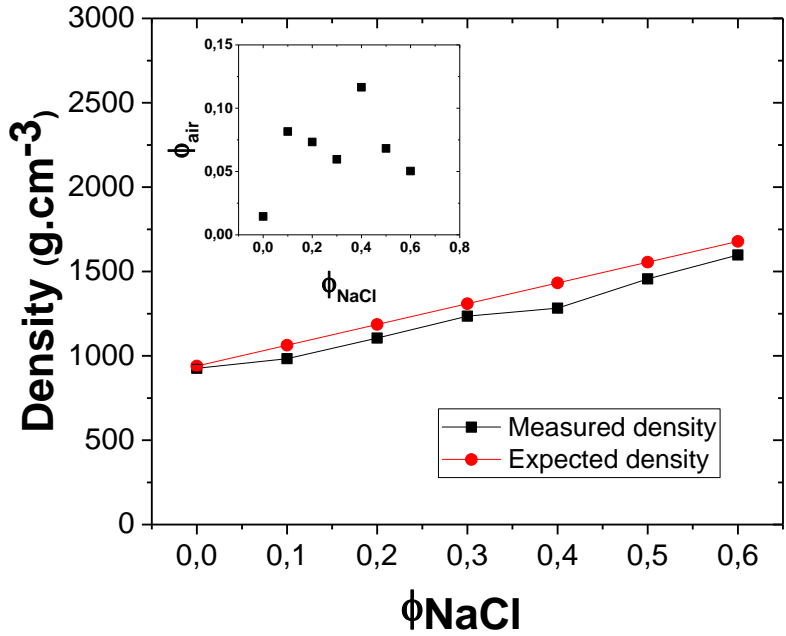

Figure 1: Comparison between the density of the samples measured with the principle of Archimedes and the expected density if no air was present in the cubes. Apart from the sample with $\phi_{NaCl} = 0.4$, the quantity of air trapped in the cubes is located between 5 and 10%.

### 3.2) Micro indentation test: the Oliver Pharr method

The so-called Oliver Pharr method (14) (15) has been developed to measure hardness in ideal elasto-plastic materials from indentation measurements. It relies on measuring the load response P (in Pa) to a loading and unloading strain-controlled ramp. Since only elastic/reversible forces are recording during the unloading step, one can deduce the Elastic modulus E of the material from this curve, based on the indentation h, the contact area of the probe A (m²) and the slope of the upper portion $S = \frac{dP}{dh}(h = h_{max})$ of the unloading curve at the initial stage of the unloading with the relation:

$$S = \beta \frac{2}{\sqrt{\pi}} E \sqrt{A} \tag{2}$$

Although the method has been originally developed for specific shaped probe, the method was proven to be valid for any kind of axisymmetric indenter such as cone or spheres (12). In the rest of the study, we will use a sphere as the indenting probe. It has also been proved that the Oliver-Pharr method is also valid for visco-elastic materials (22). The indentation test has been performed on 5 samples for each different $\phi_{NaCl}$. The indentation sphere has a diameter of 6.35mm, the loading speed is 0.1mm.s⁻¹ and the unloading speed is 1mm.s⁻¹. Between the loading and the unloading steps, the indenter rests 300s so that all the viscous relaxation phenomena have the time to occur

Figure 2 shows the loading and unloading curves of a few samples in function of $\phi_{NaCl}$. In term of maximum force at the end of the loading phase, an initial weakening between the samples made of pure fat and the samples with 10% of salt can be witnessed, most likely due to the introduction of air with the solid particles. Figure 2(b) focuses on the beginning of the unloading curve, where the slope is extracted to deduce the young modulus E thanks to equation (1) (inset). For samples with very low $\phi_{NaCl}$, visco-plastic deformations are very high, and therefore, the unloading curves diverge quickly from the linear shape. The indentation range where the curve can be assimilated to a slope is very low and the precision of the measurement is thus impacted. Those measurements are probably incorrect: When looking at the moduli extracted from the slopes, one can see that the difference between the lowest and the highest values are very low, around 50%. However, when qualitatively comparing the loading-curves, one can see that increasing $\phi_{NaCl}$ has a big impact on the force necessary to reach the maximum indentation (from 2 to 10N), but also on the reversibility of the loading-unloading process: The young modulus should therefore increase much more than the mere 50% observed in the results.

A possible explanation for the failure of this measurement is certainly the very reduced linear domain of our samples at low $\phi_{NaCl}$. But another reason might explain why the Oliver Pharr method fails to describe the Young Modulus of our materials: the time necessary for the probe to completely unload has to be ten times faster than the characteristic relaxation time of our samples (20). Since the texture analyzer maximum velocity is 1mm.s$^{-1}$ , this makes total unloading time of the probe equal to 0.5s. If the characteristic relaxation times scales are lower than 50ms, the micro indenter will fail to record the whole relaxation process. As we will see in the next section, some of the characteristic time scales of the material are indeed much smaller than that, therefore invalidating this method.

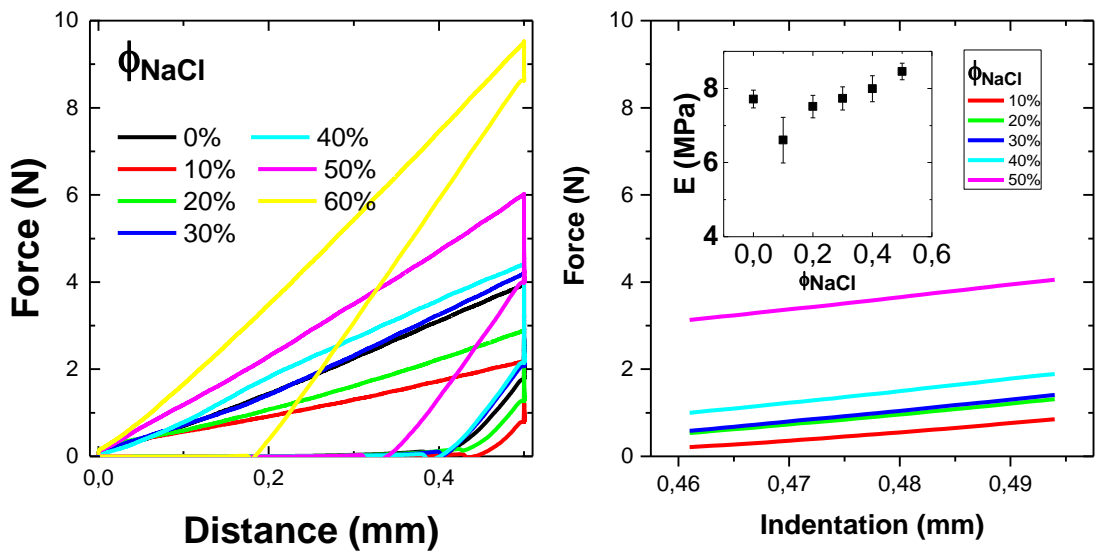

Figure 2. Left: typical results obtained when applying the Oliver-Pharr method on binary samples for different $\phi_{NaCl}$. The upper part of each curve represents the loading process and the lower part of each curve the unloading process.
Right : The curves have been trimmed to show only the beginning of the unloading process from which we can derive the elastic modulus E thanks to equation (1). The moduli are plotted in the inset as a function of $\phi_{NaCl}$.

### 3.3) Micro indentation test: the stress relaxation method

Another way to characterize the samples with micro indentation is to apply an instantaneous indentation and to measure the force relaxation over time. From these measurements, it is possible to check if we are in the linear regime, to extract characteristic time scale, and to simulate various load protocol to predict the behavior of the samples. Of course, those expectations are valid only in the linear regime and cannot predict fracture for instance (16). A few studies (12) (20) (13). focused on the theoretical behavior of visco-elastic samples under indentation and derived predictive formulas depending of the shape of the probe and the characteristic of the sample according to a simple linear model.

The first step of the protocol is to check if our samples have the same behavior regardless of the indentation depth of the probe. We indented different samples with varying indentation depth for $\phi_{NaCl} = 0.6$ and recorded the relaxation for 100s. We then tried to rescale all the curves to verify if they can superimpose into a single master curve. If this is the case, we are then in the linear regime and linear models can be used to describe our materials (until non-linear phenomena such as fracture happen).

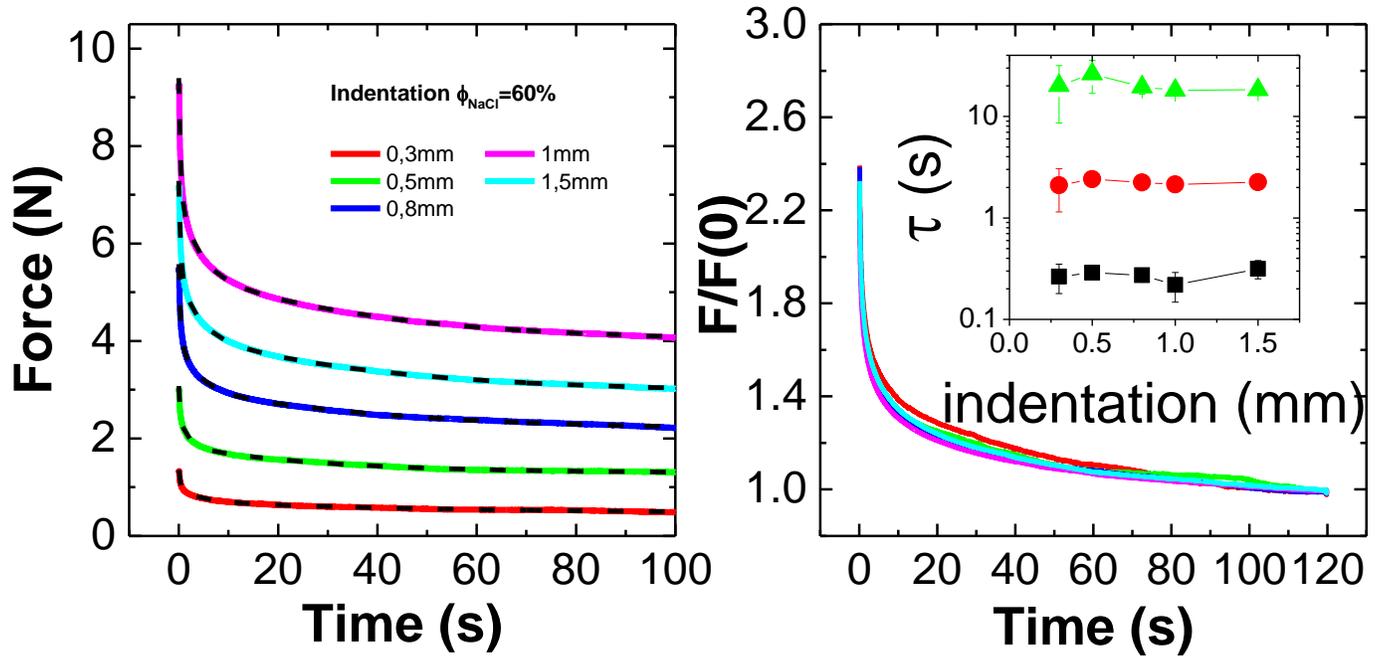

Figure 3: (Left) Relaxation of the force applied to the samples after immediately reaching a given indentation for $\phi_{NaCl} = 0.6$. The coloured lines represent the experimental results, the dashed lines represent a fit obtained from the sum of three exponential decays. (right): The relaxations curved are rescaled and collapse into a single one, showing that we are in the linear regime. The time constants obtained from the fits are plotted in the inset and are also insensitive to the indentation depth.

Figure (3) displays the data for different indentation depths, with their absolute value (a) and then rescaled. All the data collapse into a master curve, showing that the linear regime is still valid. We tried to fit the curves with a single exponential decay but failed to get good results, implying than more than one time constant is involved in the relaxation process. Fitting the curves with a sum of three exponential data (equation 3) give satisfactory results ($R^2$>0.998) for all the results present in the study.

$$F(T) = B + A_1.e^{-\frac{t}{\tau_1}} + A_2.e^{-\frac{t}{\tau_2}} + A_3.e^{-\frac{t}{\tau_3}} \tag{3}$$

In the inset of fig 3b, the time constants of the exponential decays are plotted. The error bars represent the standard deviation over 5 samples. The time constants are also insensitive to the change of indentation depth, hinting that we indeed remain in the linear regime.

We can now focus on the results of the relaxation forces on samples for different values of $\phi_{NaCl}$. Results are shown on fig 4. Panel (a) shows typical relaxation curves obtained for different volume fractions. We can already make a few observations here. First, at low $\phi_{NaCl}$, we can retrieve the same trend than what was observed with the Oliver-Pharr experiments: The force measured for pure fat samples are higher than for samples with a low $\phi_{NaCl}$, once again most likely due to the introduction of air in the cube alongside the solid particle. Interestingly enough, the relaxation rate $\frac{F(t=\infty)}{F(t=0)}$ of the pure fat sample is higher than for the other curves: The curves for the different samples cross each other. When increasing $\phi_{NaCl}$, the force required to deform the samples increases then consequently, with roughly a factor 10 difference between the softer and the harder samples.

Because of this big difference in relaxation rate between all the samples, it is impossible to linearly rescale the curves. Such an attempt has been made on panel b, but completely failed. This is understandable since the pure fat samples lose 80% of their original force over time, whereas the samples with 60% of salt lose only 20% of it. We plotted the value of the force F on the steady state regime versus the $\phi_{NaCl}$ on panel c. In the inset, we rescaled these numbers by the value of the force recorded at t=0. The decrease in steady state normal force observed for absolute value disappears once the rescaling is complete: The ratio $\frac{F(t=\infty)}{F(t=0)}$ increases monotonously. Despite being "weakened" by the introduction of air, the samples become more and more elastic with the introduction of solid particle. The last panel shows the values of the time constants obtained from the fits. The values are rather insensitive to $\phi_{NaCl}$, only their relative contribution compared to the non-vanishing part varies. We can observe a slight raise of the values at very high $\phi_{NaCl}$ but it is rather hard to conclude if this change is significant or not.

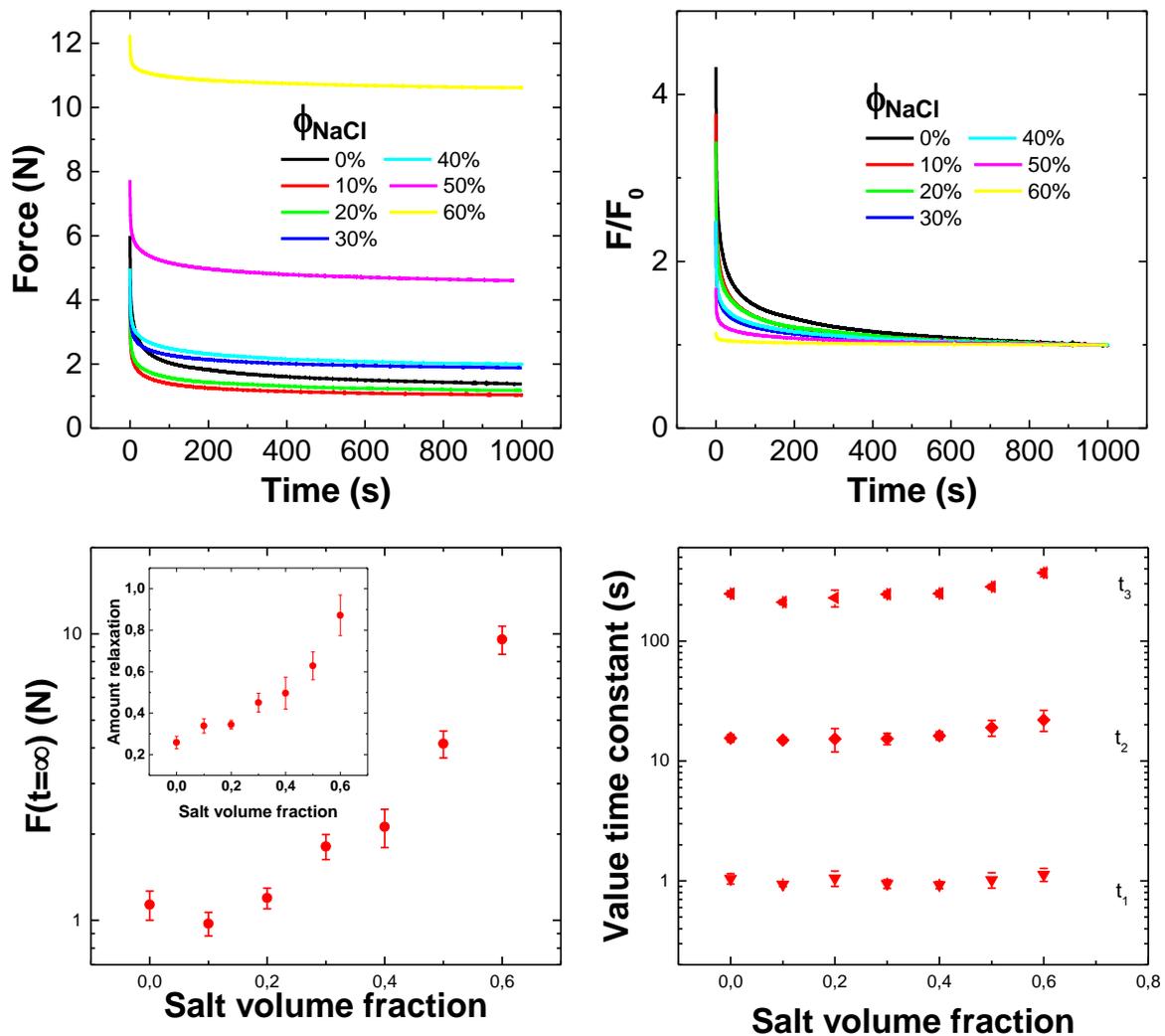

Figure 4. (Top left). Typical force-relaxation curve for varying $\phi_{NaCl}$. (Top right) An attempt to rescale the curves have similarly to figure 3. Has been made but because of the different dissipation rate depending on $\phi_{NaCl}$, such rescaling is impossible. (Bottom left): the average value of the relaxation force curve on the steady state regime. Following the initial decrease, the value then raises monotonously. Inset: the values have been rescaled by the initial value of the force at t=0. This time, the initial weakening disappears. (bottom right): Value of the time constants obtained from the fits versus $\phi_{NaCl}$.

## 3.4) Rheological model

From those relaxation measurements, it is possible to derive a simple rheological model. Such a model can be useful to simulate different kind of loading procedures and predict the force necessary to enforce them (and vice versa), if we remain in the linear domain. The relationship between the relaxation curve measured and the values of the springs and dashpots of a classical three model elements for a spherical indenter is known from previous theoretical work (12) (20). Since our relaxation curves are fitted by a sum of three exponential decays, a combination of three spring and dashpots is necessary to simulate our curves. Since we are in the linear regime, the principle of superposition is valid (16). Fig 5 shows the rheological model used to reproduce experimental results. In this model, three 3-elements models are in parallel, with the same spring $E_4$ for each model, since it represents the steady state forces of the model. The relaxation curves are expected to follow the relation:

$$F(t) = \frac{16}{9}\sqrt{\delta_0^3 R} E_4 \sum_{i=1}^{3}\left(\frac{E_4}{E_4 + E_i} e^{\left(\frac{E_4+E_i}{3\eta_i}\right)t} + \frac{E_i}{E_4 + E_i}\right) = \frac{4}{3}\sqrt{\delta_0^3 R}.G(t) \qquad (4)$$

In our model, 7 elements have to be determined: $E_{1,2,3,4}$ and $\eta_{1,2,3}$. In our relaxation fits, we have 7 parameters, $A_{1,2,3}$, $\tau_{1,2,3}$ and B. The value of the different elements of the models can therefore be calculated with a simple identification between the equations (3) and (4).

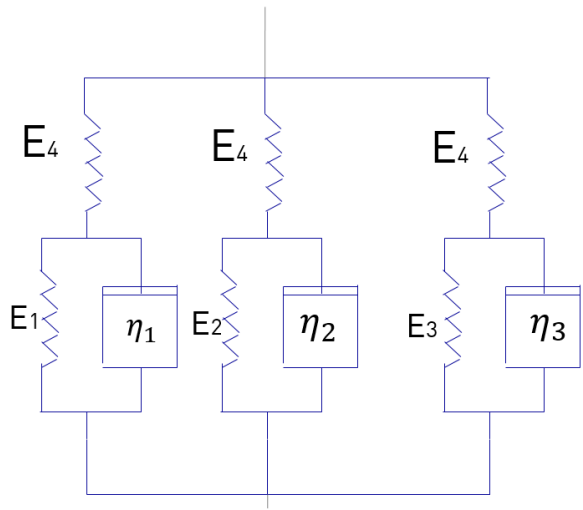

Figure 5: Rheological model used to simulate the deformation behavior of our binary samples. The springs represent the elastic and reversible deformation and the dashpots represent the viscous events. Since our relaxation curves are fitted by a sum of three exponentials, three combinations of springs and dashpots are necessary.

A Matlab script was written to identify the elements of the model according to the relaxation curves, and the value of the parameters was estimated with the least square method. In this script, the 7 fitting parameters, $A_{1,2,3}$, $\tau_{1,2,3}$ and B are calculated from the relaxation curves and then identified with equation (3), to give the parameters of the model. Values of the parameters for varying $\phi_{NaCl}$ are plotted on figure (6).

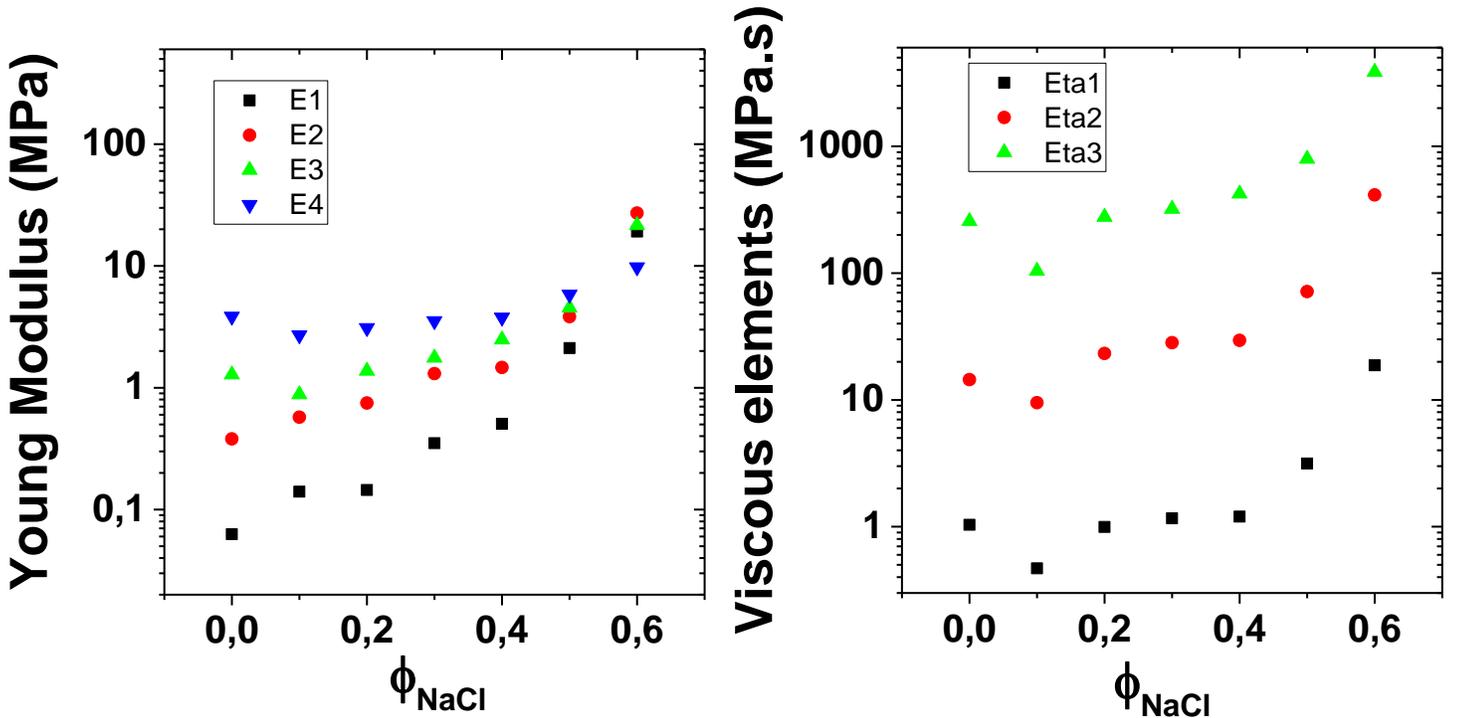

Figure 6: Values of the different elements of the rheological model obtained by identification of the fits with equation (3) versus $\phi_{NaCl}$. All the values increase monotonously except for the first points.

As can be seen, the value of the parameters increases with $\phi_{NaCl}$. But we can also remark that the elements associated with the highest time constant (E3,$\eta$3) does not vary as much as the ones associated with the shortest time constant. This explains why, in our Oliver Pharr measurements, Young moduli results did not vary that much with $\phi_{NaCl}$ : the unloading speed was too slow to capture what was happening at the smallest time scale, where most of the change was happening.

From the value of these parameters, it is possible to derive some information such as the complex modulus of the material. Indeed, in the linear domain the complex modulus can be calculated with the following formula:

$$G^*(\omega) = \sum_{i=1}^{3} \frac{1}{\frac{1}{E_4} + \frac{1}{E_i + j\omega\eta_i}} \tag{5}$$

Figure 7 shows the complex modulus derived from the rheological model for all the different $\phi_{NaCl}$. From these curves, we can plot the phase angle (showing the competition between the solid and the viscous contribution of the material) as well as the complex modulus for a given frequency as a function of $\phi_{NaCl}$. Figure 7 b and c show those data for the frequency 0.5Hz, where the viscous contribution is the highest for most of the volume fraction. Interestingly enough, the complex modulus G* can be fitted by the classical empirical Krieger-Dougherty relation $G = G_0 * (1 - \frac{\phi}{\phi_{max}})^{-n.\phi_{max}}$ with $\phi_{max}$, the maximum volume fraction is found equal to $\phi_{max} = 0.69$. On the other hand, the phase angle versus $\phi_{NaCl}$ curve can be fitted by a line, which intersects the y=0 axis (where the product is completely solid) at $\phi_{max} = 0.68$.

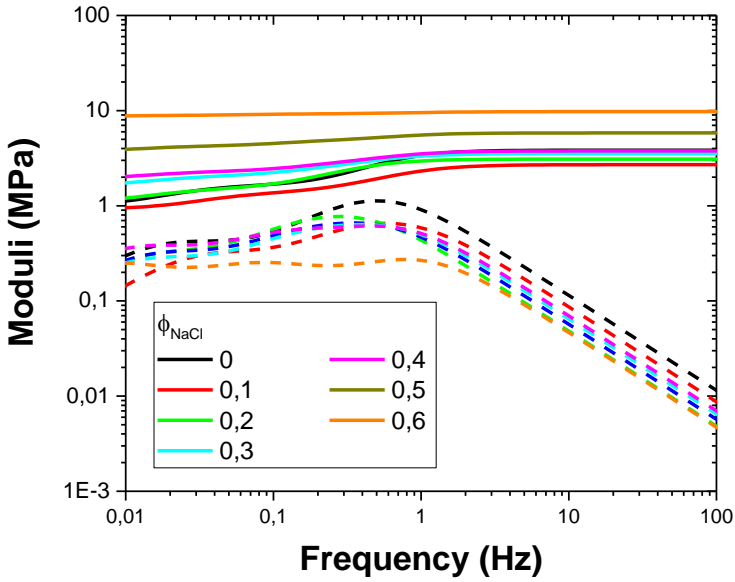
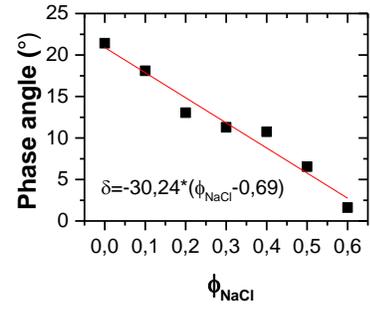
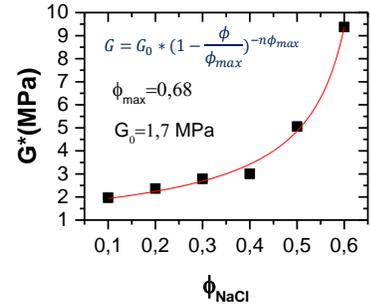

Figure 7: (Left): Value of the storage and loss moduli versus frequency derived from the rheological model for various $\phi_{NaCl}$. (top right) Phase angle versus $\phi_{NaCl}$ at f=0.5Hz. The red line is a linear fit which meets the axis y=0 for $\phi_{NaCl} = 0.69$. (Bottom right): Value of the storage modulus G' as a function of $\phi_{NaCl}$ at f=0.5Hz. The red curve is a fit obtained from the classic empirical Krieger Dougherty relation with $\phi_{max} = 0.69$

The main interest of this model is that from the results obtained from the force relaxation with an instantaneous creep response, one can predict the force response of the material from any kind of loading behavior, as long as we remain in the linear regime. Indeed, if G(t) in equation 3 is known, then one can predict the shape of the force for a given arbitrary displacement $\delta(t)$ with the formula:

$$F(t) = \int_0^t (2\sqrt{R\delta(t)}G(t-\tau)d(\delta(\tau)) \qquad (6)$$

To test the validity of this theory and of the hypothesis of the linear model, we simulated with Matlab the theoretical loading curve that our binary samples would have when subjected to a linear load $\delta(t) = b.t$ with b being the loading velocity of the probe. We then applied experimentally the same loading profile to our cubes with a Texture analyser (loading speed 10um.s$^{-1}$, spherical probe of 10mm diameter) and we compared the experimental results with the predictions obtained from the rheological model. The results for two different $\phi_{NaCl}$ (0 and 50%) are shown in figure 8. The results are in very good agreement with each other, until non-linear effects such as weakening or fracture occurs. This is also the case for the other $\phi_{NaCl}$ (not displayed).

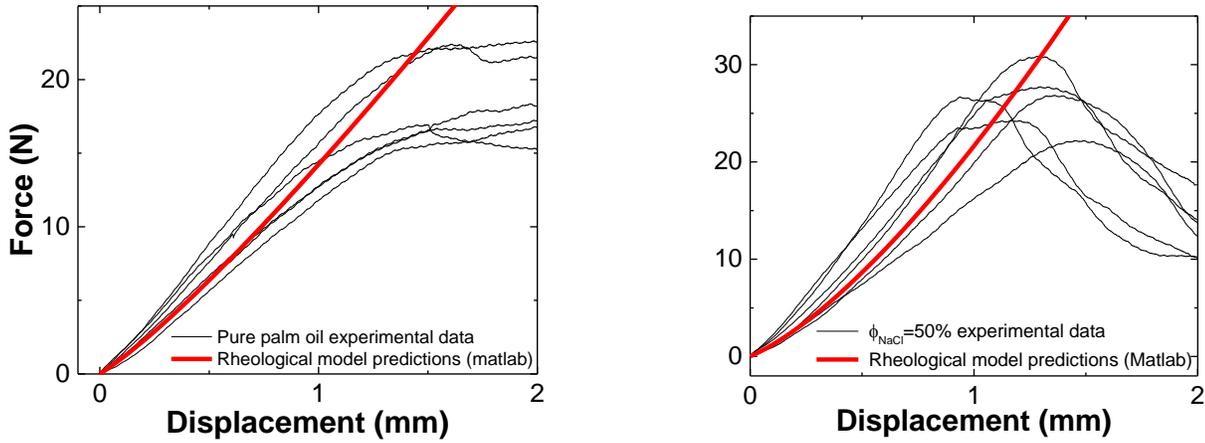

Figure 8: Experimental force versus displacements loading curves obtained with a linear indentation (10um.s$^{-1}$) of a spherical probe (diameter : 10mm) for pure palm oil samples (left) and binary systems with $\phi_{NaCl} = 0.5$. The red curves are the predictions obtained by the rheological model derived from the force relaxation measurements.

**4) Conclusion**

In this paper, we studied the physical properties of semi solid heterogeneous mixes made of crystallized fat matrix with solid inclusions in the form of sodium chloride crystals. Given the general shape and hardness of our samples, oscillatory rheology was not suitable to measure their flowing properties. We therefore explored a few protocols with a texture analyzer. The so-called Oliver-Pharr method, where the force-displacement curve of the materials is measured during loading and unloading proved to be ineffective, because of the speed limitation of our device. Measuring the stress relaxation for a long time proved to be much more efficient. From those measurements, we set up a very simple linear rheological model allowing us to estimate the Storage and loss moduli from our samples. We showed that replacing the fat by sodium chloride crystals make the material harder and increases the relative contribution from the elastic forces (therefore, diminishing the plastic deformations). The evolution of the complex modulus follows the classical empirical Krieger-Dougherty law, ultimately diverging at $\phi_{NaCl} = 0.68$. The linear rheological model was also able to predict from the stress relaxation data the force-displacement relation curve for a linear indentation (in the small deformation limit).

The experimental values obtained by this indicate that salt crystals have an active contribution on the small-deformation rheological behavior of fat-based systems. Moreover, we find no evidence that the rheological properties of the continuous fat phase are affected by the presence of the salt crystals up to $\phi_{NaCl} = 0.6$.

We can hope in the future to measure the properties of cubes with a tertiary component in it. More specifically, we are very interested in adding a small amount of water in the cubes, in order to create capillary bridges. It has been proved in hydrophobic suspensions that the addition of water changes drastically the structure, increasing the storage modulus of the suspension of a few decades. We want to see if such a structure could be achieved in a material where the continuous part is not completely liquid, but viscoelastic.